\documentclass[12pt]{article}
\usepackage{makeidx}
\usepackage{multirow}
\usepackage{multicol}
\usepackage[dvipsnames,svgnames,table]{xcolor}
\usepackage{graphicx}
\usepackage{epstopdf}
\usepackage{amsmath,amssymb,amsfonts,amsthm}
\usepackage{ulem}
\usepackage{hyperref}
\usepackage{mathrsfs}
\author{Yuri Kamyshkov}
\title{}
\usepackage[paperwidth=612pt,paperheight=792pt,top=72pt,right=72pt,bottom=72pt,left=72pt]{geometry}

\makeatletter
	{\par\setlength{\parindent
}{#3}
	\setlength{\leftmargin}{#1}       \setlength{\rightmargin}{#2}%
	\advance\linewidth -\leftmargin       \advance\linewidth -\rightmargin%
	\advance\@totalleftmargin\leftmargin  \@setpar{{\@@par}}%
	\parshape 1\@totalleftmargin \linewidth\ignorespaces}{\par}%
\makeatother 

\begin{document}

\begin{center}
\textbf{{\large Neutron Disappearance and Regeneration from Mirror State}}
\end{center}

\begin{center}
\textit{Zurab Berezhiani$^{1}$, Matthew Frost$^{2}$, Yuri Kamyshkov$^{2}$, Ben Rybolt$^{2}$, Louis Varriano$^{2}$}
\end{center}

\begin{center}
\textit{{\small $^{1}$Dipartimento di Fisica e Chimica, Universit\`a di L'Aquila, 67010 Coppito AQ;\\
INFN, Laboratori Nazionali del Gran Sasso, 67010 Assergi AQ, Italy;\\
$^{2}$Department of Physics, University of Tennessee, Knoxville, TN 37996-1200, USA}}
\end{center}

\date{\today}

\begin{abstract}
\small
The purpose of this paper is to demonstrate that if the transformation of a neutron to a mirror neutron exists with an oscillation time of the order of ten seconds, it can be detected in a rather simple disappearance and/or regeneration type experiment with an intense beam of cold neutrons. In the presence of a conjectural mirror magnetic field of unknown magnitude and direction, the resonance transformation conditions can be found by scanning the magnitude of the ordinary magnetic field in the range e.g. $\pm 100 \mu$T.
Magnetic field is assumed to be uniform along the path of neutron beam. If the transformation effect exists within this range, the direction and possible time variation of the mirror magnetic field can be determined with additional dedicated measurements.   
\end{abstract}

\section{Introduction}

Thus far the entire notion of the existence of Dark Matter (DM) has been based on the manifestations of its gravitational effects at cosmological scales, ranging from the observations of galactic rotation curves to the measurements of the anisotropy of the CMBR \cite{Olive:2016xmw, Freese:2017idy}.
Many different theoretical models of Dark Matter ~\cite{Olive:2016xmw, Freese:2017idy} 
suffer from a lack of direct experimental measurements that could reveal the nature of DM  
in interactions other than gravitational. 
Several direct Dark Matter search experiments ~\cite{Freese:2017idy} have created a rather controversial situation, 
from one side excluding Weakly Interacting Massive Particles (WIMPs) in the range of masses $\gtrsim$~6 GeV with the weak coupling strength to the target particles of Ordinary Matter (OM), and from another side claiming signals of Dark Matter detection in a lower mass range than expected by the popular supersymmetric models \cite{Munoz:2017ezd}. 
On the other hand, experiments at the Large Hadron Collider so far have not been able to find WIMP particles  
that could be candidates for a single-component DM ~\cite{Freese:2017idy}. 
The general difficulty in identifying the nature of Dark Matter is the lack of non-controversial and reproducible effects of non-gravitational interactions between the DM and OM particles. 

In this paper we will discuss a new, simple particle physics experiment with neutrons that could test one of 
the DM  models of specific interest. In this model, DM (in whole or at least in some fraction) consists of Mirror Matter (MM) from a hidden gauge sector that represents an exact copy of the ordinary particle sector. 
The concept of mirror fermions was introduced in 1956 as a way to restore parity in a seminal paper by Lee and Yang, 
while in 1966 Kobzarev, Okun and Pomeranchuk  showed that mirror fermions cannot have 
common strong, weak and electromagnetic interactions with ordinary fermions and 
thus the former can exist in the form of an independent parallel sector which they coined as the mirror world \cite{Mirror}. 
From todays' view, 
once ordinary matter and its interactions are described by the Standard Model (SM), then the mirror 
particle content and interactions are  described by another copy of (SM$'$). 
It can be equivalently viewed, in the context of extra dimensions, as OM and MM are localized on two parallel branes with corresponding SM and SM$'$ self-interactions, while the gravity propagating through the bulk interacts with both types of 
matter.   Thus, MM and OM interact gravitationally, providing observable manifestations of DM effects, 
but they have no common SM and SM$'$ interactions; in fact, renormalizable interaction terms between particles 
of two sectors are forbidden by the particle content of the Standard Model, in the same way as renormalizable 
interactions   violating baryon or lepton numbers. 
However, in difference from cold dark matter candidates, MM is self-interacting and dissipative matter in the same sense as OM. 
Once MM is adopted as a framework for DM, then the properties of MM with respect to particle content and interactions are well known by the model requirement of SM $\equiv$ SM$'$. On the other hand, the cosmological genesis and galactic evolution might be different between MM and OM  due to differences in abundance and temperature of the two sectors and the obvious lack of observational details of the mirror sector. Many aspects of the MM cosmology were discussed and addressed in the  literature \cite{Berezhiani:2000gw}.      
Although these aspects are not broadly assimilated, the mirror matter hypothesis  remains a viable option for DM. 
A detailed discussion of the viability of the MM model is not in the scope of this paper. The interested reader can learn these ideas e.g. from  reviews \cite{Okun:2006eb}. 

Important for this paper are two aspects of MM. First, it was conjectured that the neutral particles of the OM and MM sectors, elementary like neutrinos or composite like neutrons,  can be mixed, providing an interaction portal between the two sectors. 
Namely, the kinetic mixing of ordinary and mirror photons, first discussed in \cite{Holdom:1985ag}, 
which makes mirror matter effectively mini-charged with respect to the ordinary electromagnetic forces.  
Though the kinetic mixing is severely restricted by cosmological limits \cite{Lepidi}, 
it can lead to experimentally observable effects, such as positronium oscillation into mirror positronium \cite{Gninenko}, and can also 
provide an interesting portal for the direct detection of mirror dark matter consisting dominantly of mirror helium and hydrogen  and a smaller fraction of heavier mirror nuclei like carbon, oxygen etc. \cite{DAMA}. 
In addition, ordinary and mirror particles can also share some hypothetical forces, interacting e.g. with gauge bosons of common family symmetry \cite{SU3} or common  $B-L$ symmetry \cite{B-L}.   
The lepton and baryon number violating interactions between OM and MM are of specific interest since 
they could lead to the co-genesis of baryon asymmetries in both sectors and shed the light on the near coincidence 
of the baryon and dark matter fractions in the Universe \cite{BB}.  Such interactions, at low energies, would induce lepton or 
baryon number violating effects such as the ordinary (active) neutrino oscillations into mirror (sterile) neutrinos \cite{neutrinos} 
and neutron oscillations into mirror neutrons \cite{Berezhiani:2005hv},  resembling and perhaps related to the
more familiar but still not observed phenomenon of neutron--antineutron oscillation \cite{nnbar}. 
 
Particularly interesting for us is the effect of neutron to mirror neutron transformation ($n \rightarrow n'$) first theoretically considered in \cite{Berezhiani:2005hv}. It was noticed in this paper that a $n \rightarrow n'$ oscillation time
larger than $\tau \sim 1$~s cannot be excluded either by the existing at that time experimental data or by cosmological and astrophysical bounds. The reason why such a fast oscillation (in fact, faster than the neutron decay), which leads to the violation of baryon (neutron) number could be unnoticed in experiments 
is the presence of the Earth magnetic field, which was never compensated for in the experiments measuring the neutron lifetime. 
In fact, in the absence of matter and with a compensated magnetic field half of the free neutrons created in experiments would have to disappear within seconds.  

However, it was recognized later that a mirror magnetic field $B'$, as a natural component of the MM model, might be present but unnoticed in terrestrial experiments, possibly being of galactic origin or due to the accumulation of MM inside the Earth 
\cite{Berezhiani:2008bc}. This is a second aspect of MM that is important for this paper. In particular, the photon -- mirror photon kinetic mixing \cite{Holdom:1985ag}   could induce large galactic magnetic fields, both ordinary and mirror, via the electron drag mechanism \cite{BDT}.  
On the other hand, the same interaction portal could lead to the accumulation of some small amount of mirror matter
 in the Earth, which could be sufficient  for the induction of a mirror magnetic field up to few Gauss via the mirror electron 
drag due to rotation of the Earth, the mechanism discussed in Ref. \cite{BDT}.  
In this way, the mirror magnetic field at the Earth might have a magnitude of the same order as the Earth's own magnetic field.  
In addition, its orientation relative to ordinary magnetic field can be arbitrary and might also be variable in time \cite{Berezhiani:2008bc}.

\section{Previous searches for UCN disappearance}

The idea of a possible neutron to mirror neutron transformation \cite{Berezhiani:2005hv} and following 
paper discussing experimental sensitivities for its search \cite{Pokot} 
has stimulated several experimental searches \cite{Ban,Serebrov1,Serebrov2,Altarev}, the results of which were adopted by the Particle Data Group \cite{Olive:2016xmw}.  
These experiments were performed with Ultra-Cold Neutrons (UCN), where the effect of $n \rightarrow n'$ transformation could lead to the unaccounted disappearance of the neutrons stored in the UCN trap. In the presence of Earth magnetic field $B$, the neutron with its magnetic moment $\mu$ 
will change its energy level by $\mu B$, while the mirror neutron   will not interact with ordinary magnetic field $B$. 
Thus, the Zeeman split of the energy levels would suppress the $n \rightarrow n'$ transformation. The first experiments were thought to require shielding of the Earth's magnetic field, due to their assumption that no mirror magnetic field could exist at the Earth.
In this way, under the hypothesis that the mirror magnetic field at the Earth is vanishing, the limit $\tau > 414$ s 
was  obtained at 90 $\%$ C.L. \cite{Serebrov1,Olive:2016xmw}. 

In the later experiments the existence of a non-zero mirror magnetic field $\bf B'$ was admitted into the hypothesis. In this case, the detection of a disappearance signal would require the tuning of the ordinary laboratory magnetic field $\bf B$ and would have a resonance character in the transformation probability if the magnitude of the field $B$ were the same as that of the mirror field $B'$. Additionally, the transformation effect would be enhanced when the directions of both fields would be aligned and would be minimal when directions of $\bf B$ and $\bf B'$ were opposite. 

The probability of $n \rightarrow n'$ transformation in the absence of any fields can be described, as for any two-level system, by the proper oscillation time $\tau=\hbar / \epsilon$, where $\epsilon \lesssim 10^{-16}$ eV  ~\cite{Berezhiani:2005hv} is a small mass mixing $n-n'$. In the presence of two different magnetic fields $\bf B$ and $\bf B'$, time evolution of the probability of $n \rightarrow n'$ transformation was described in ~\cite{Berezhiani:2008bc, Berezhiani:2012rq} by exact solution of  Schr\"{o}dinger equation with a $4 \times 4$ Hamiltonian describing both the neutron and mirror neutron polarizations and the mixing between the neutron and mirror neutron components. Since the magnetic moment of neutron (or mirror neutron) $\mu \simeq - 6 \times 10^{-12}$ eV/G, the energy split in milli-Gauss magnetic fields $\vert \mu B - \mu' B' \vert > \epsilon$ 
and the probability of $n\to n'$ (or $n'\to n$) transformation after free flight time $t$ can be simplified ~\cite{Berezhiani:2012rq} to
\begin{equation}
P_B(t) = \mathcal{P}_B(t) + \mathcal{D}_B(t) \cdot \cos\beta
\end{equation}
where $\beta$ is the angle between the magnetic field vectors $\bf{B}$ and $\bf{B'}$ and
\begin{align}
\mathcal{P}_B(t) = \frac{\sin^2[(\omega-\omega')t]}{2\tau^2(\omega-\omega')^2} + \frac{\sin^2[(\omega+\omega')t]}{2\tau^2(\omega+\omega')^2} \nonumber \\ \\
\mathcal{D}_B(t) = \frac{\sin^2[(\omega-\omega')t]}{2\tau^2(\omega-\omega')^2} - \frac{\sin^2[(\omega+\omega')t]}{2\tau^2(\omega+\omega')^2} \nonumber
\end{align}
with $\omega = \frac{1}{2}|\mu B|$ and $\omega' = \frac{1}{2}|\mu B'|$, where $\mu$ is the magnetic moment of the neutron and mirror neutron, assumed to be identical ~\cite{Berezhiani:2012rq}. We see that both terms $\mathcal{P}_B(t)$ and $\mathcal{D}_B(t)$ in the equation (2) have a part with resonance behavior when the magnitude of the magnetic field $B$ is close to the unknown magnitude of mirror magnetic field $B'$. Resonance occurs even if the
directions of the $\bf B$ and $\bf B'$ vectors do not coincide, but the probability is maximum when $\cos \beta =1$ and minimum when $\cos \beta = -1$.

As was mentioned above, all previously published experiments looking for the disappearance of $n \rightarrow n'$ were performed with UCN vacuum traps. A relatively small number of UCN were stored in the traps by reflections off the walls coated by material with a high nuclear Fermi-potential. The typical velocity of UCN was a few m/s and the time between collisions was typically around 0.1 s. A controlled, intentionally-uniform magnetic field was applied to the UCN volume and was varied during the experiments. Neutrons that transformed to the mirror state would have become sterile to interactions with OM nuclei in the walls and would have escaped the trap, thus leading to an increase of the disappearance rate.

In the experiment \cite{Ban}, the $n \rightarrow n'$ effect was sought for by studying the variation of the storage time of UCN in the trap with reflecting walls by switching on and off the external magnetic field $B$. In ``off'' mode, the magnetic field in the trap was close to zero (the Earth magnetic field was compensated down to $\lesssim$ 100 nT). No disappearance signal was found corresponding to the oscillation time limit $\textgreater$ 103 s (95$\%$CL).

In a more sophisticated UCN experiment \cite{Serebrov1, Serebrov2} under similar magnetic field $B$ suppression, this limit (assuming no mirror magnetic field) was improved to an oscillation time of $\tau >  414$ s (at 90$\%$CL). These limits, however, are not valid if one assumes that the mirror magnetic fields are present. 

The experimental data of the latter experiment \cite{Serebrov1, Serebrov2} were made available and re-analyzed by the authors of Ref. \cite{Berezhiani:2012rq} with a more advanced theoretical oscillation description from \cite{Berezhiani:2008bc}. Additionally, in this re-analysis some data runs corresponding to unstable reactor flux conditions, as indicated by the system of two monitoring counters, were excluded. The neutron storage time in the trap was measured multiple times with variation of the laboratory magnetic field that was regularly alternated in the vertical direction with magnitude $\pm 20 \mu T$.
From these data the asymmetry parameter from up and down variation of the direction of the magnetic field was calculated and the result has shown a peculiar anomaly with statistical $5 \sigma$ deviation from the expected zero asymmetry for non-polarized neutrons in the Be-coated trap.
In the absence of consistent alternative interpretations, the asymmetry could be explained as an effect of the variation of the angle between $\bf B$ and unknown $\bf B'$ resulting in a different degree of suppression of the $n \rightarrow n'$ transformation probability. The model  \cite{Berezhiani:2012rq} would describe the observed asymmetry by the following parameters: an oscillation time on the order of $10$~s and the magnitude of the mirror magnetic field $\sim 10$~$\mu$T. These parameters, however, should be considered as rather approximate for several reasons. First, the angle between $B$ and $B'$ remains unknown; and second,  the effect of possible non-uniformity of magnetic field $B$ was not taken into account.

In still another experiment \cite{Altarev}, an attempt was made to scan the magnitude of the vertical magnetic field in the range 0 to $\pm$ 12.5 $\mu$T at several values of $B$ with increments of 2.5 $\mu$T.
The analysis here had a statistically insignificant indication of the transformation effect with oscillation time 21.9 s (other fit parameters were $B'=11.4 \mu$T and the angle between vertical up $B$ and $B'$ $\sim 25$ degree). The data of this experiment were also interpreted for exclusion at 95$\%CL$ the oscillation time 
$< 12 s$ within the range of studied magnitudes of field $B$. 
A drawback of this experiment was the rather large incremental step in the variation of the magnitude of magnetic field $B$. This step was more than two times wider than the width of the resonance in $|B-B'|$.

Thus, the results of all these experiments are controversial. However, they do not exclude unambiguously the possible existence of the effect in the range of oscillation time $\sim$ 10 s for a mirror magnetic field with a magnitude similar to or lower than that of the Earth magnetic field.

The common difficulty in the UCN trap experiments is that the small absorption coefficients  of the reflecting walls of the trap are experimentally measured to be larger than those theoretically predicted ~\cite{Pokotilovski-2016} from the known nuclear potentials. That leaves the room for the claimed asymmetry anomaly ~\cite{Berezhiani:2012rq} explanation by some other possible still unidentified magnetic-field-sensitive effects that are not related to the mirror state transformations.  Thus, just repeating the UCN asymmetry experiment ~\cite{Serebrov1, Serebrov2, Altarev, Berezhiani:2012rq}, could produce the same anomalous signature without resolution of its nature. In this paper we consider different method of testing the $n \rightarrow n'$ transformation hypothesis and the resolution of existing experimental anomaly, the method where multiple collisions of neutrons with $OM$ walls of the trap would be excluded. 

Another existing experimental controversy should be mentioned as well in this connection: the precision neutron lifetime measured by the appearance method in the beam of cold neutrons is significantly larger than the neutron lifetime measured by the effect of disappearance in the UCN traps with the large number of wall collisions ~\cite{Olive:2016xmw}. Possible relation of this controversy to the hypothetical  $n \rightarrow n'$ transformation was mentioned e.g. in the papers ~\cite{Wietfeldt:2014rea}.

\section{Disappearance and Regeneration effects}

In this paper we are considering two possible means of observation of $n \rightarrow n'$ transformations. Both can be combined in one experimental layout using a neutron beam from a cold cryogenic moderator. The first method is the 
observation of a $\it{disappearance}$ effect of the neutrons flying in vacuum through the path of length $L_1$ in a controlled magnetic field $B$. This uniform magnetic field should be varied by steps systematically in order to find the resonance condition where the probability of neutron disappearance due to $n \rightarrow n'$ transformations will be maximal. Beam neutrons will not interact with the chamber walls except at the entrance and exit of the vacuum volume. At the end of $L_1$ flight distance the cold neutron beam will be stopped in a $^{3}He$ gaseous detector that will be thick enough to absorb $>99\%$  of slow sub-thermal neutrons and to accurately measure beam intensity for the 
detection of the disappearance effect. The disappearance effect of the order of $\sim 10^{-7}$ should be expected in the resonance peak for an oscillation time of $\tau=20$ s. Produced mirror neutrons will not interact with this absorption detector and will propagate free through the large-thickness beam dump wall behind the detector. 

Thus, mirror neutrons passing through the absorbing detector and the absorbing wall will form a beam of mirror neutrons. If a second vacuum volume with neutron flight path $L_2$ is placed behind the absorbing wall with the magnetic field controlled to the same value of resonance $B$ as in the first volume, mirror neutrons will have a similar probability to be regenerated from mirror state to ordinary state and be detected. This is the second $\it{regeneration}$ or re-appearance method for the detection of transformation $n \rightarrow n' \rightarrow n$ where detectable neutrons would re-appear beyond the absorption wall.

The method of particle regeneration is well known, e.g. for neutral kaon transformations see \cite{Pais:1955sm}. Also, in 2007, one unpublished experimental effort was presented ~\cite{Schmidt:2007} for implementation on a cold neutron beam for disappearance and regeneration measurements with a short flight path in laboratory magnetic feiled set to $B=0$ throughout the experiment.

In our paper both disappearance and regeneration of neutrons rely on the  observation of a $B$-dependent signal of oscillation. These two methods have different systematic difficulties: in the case of disappearance, the detection of a small reduction of counting rate in the beam intensity on the order of~$\sim 10^{-6} - 10^{-7}$ is a challenge due to the need to precisely chracterize the beam intensity; and for regeneration it is the requirement of a very low background count rate, that will entail an essential detector shielding effort. Both are discussed in this paper.  In Section 4 a possible configuration of disappearance/regeneration experiment will be discussed. Further details of the potential limitations of disappearance (Section 5) and regeneration (Section 6) methods. Section 7 will summarize our discussion.   

\section{Generic Configuration of Experiment}

To demonstrate the sensitivity of a search for $n' \rightarrow n$  transformations, we consider a rather generic configuration of an experiment that could be implemented at many reactors or spallation sources where intense cold neutron beams are available (Figure 1). We assume that two neutron flight paths in series, with $L_1=L_2=15$-m long vacuum tubes will be made of non-magnetic Aluminum 6061-T6 alloy. Inside these tubes, a uniform magnetic field will be created and controlled by sets of external 3-dimensional current coils and magnetic shielding. An effective neutron absorber will be placed along the beam between the first and second tubes, and a high-efficiency $^3He$ detector with appropriate shielding will be placed at the end of the second tube to detect regenerated neutrons. Figure 1 also shows a typical collimator defining the divergence of the beam and two additional $^3He$ counters: one with typical efficiency $\sim$ 1\% upstream of the first flight tube serving as the beam intensity monitor and another one for disappearance measurement with $\sim$100\% efficiency in front of the beam-stop absorber.  

\begin{figure}[t]	
\centering
\includegraphics[width=450pt, trim= 50 130 40 115, clip=true]{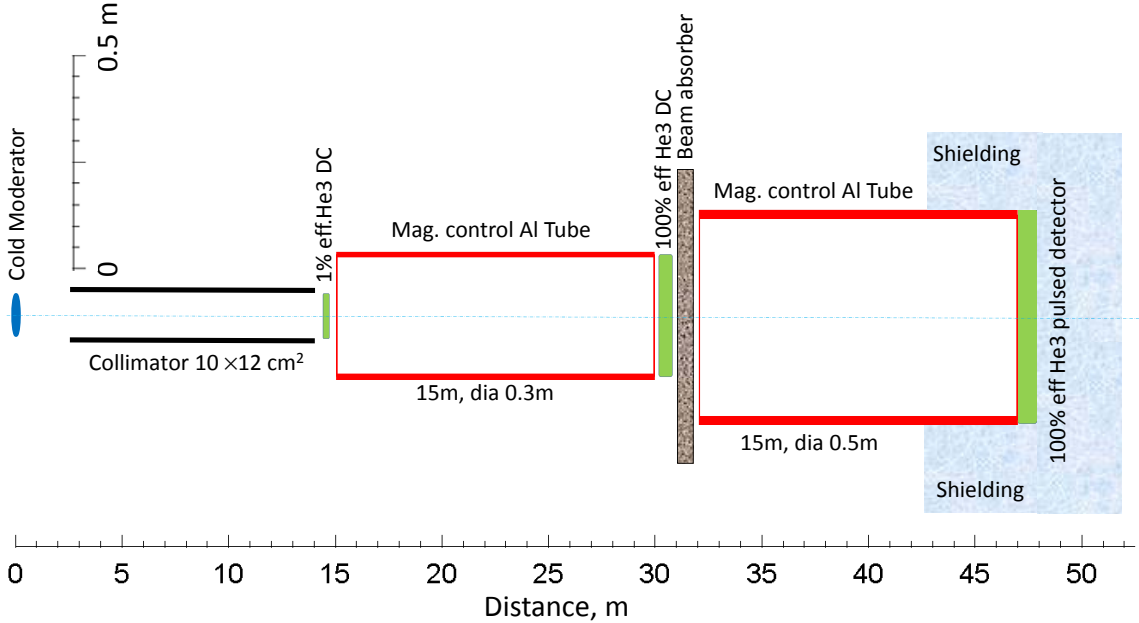}
\caption{Schematic layout of $n \rightarrow n' \rightarrow n$ regeneration experiment with a generic cold beam. Two vacuum flight tubes of 15-m length with controlled magnetic field inside the tubes are shown. First current-integrating $^3He$ beam intensity monitoring counter (A) along the beam can have efficiency e.g. 1\%. Second current-integrating beam absorption $^3He$ counter (B) with efficiency close to 100\% is for the measurement of neutron disappearance. The last $^3He$ counter (C) operating in pulse-counting mode serves for detection of regenerated neutrons. }\label{fig:1}
\end{figure}

As an example of cold neutron beam we are using the parameters of one of the neutron beams at the Spallation Neutron Source (SNS)  \cite{SNS-beam} at Oak Ridge National Laboratory. The SNS, being the latest in a series of modern and well-characterized operating sources, is a suitable candidate for such a conceptual experiment. The SNS produces neutrons via the bombardment of high-energy protons into a heavy metal (liquid mercury) target. The proton beam is pulsed at 60Hz, allowing for exploitation of the time structure to resolve thermal neutron energies to $\mu$eV resolution. While this latter capability is useful for many neutron scattering applications at SNS, a $n\to n'$ search experiment would not immediately benefit from this time structure. The proposed detector distances ensure that there is always neutron intensity available across the entire time period between pulses. The high-brilliance of the cryogenic super-critical $H_2$ moderator system provides a very intense source of sub-thermal neutrons ~\cite{SNS-beam} with extracted spectrum $S(v)$ shown in Figure 2.

\begin{figure}[t]	
	\centering
\includegraphics[width=450pt, trim=40 80 40 90
, clip=true]{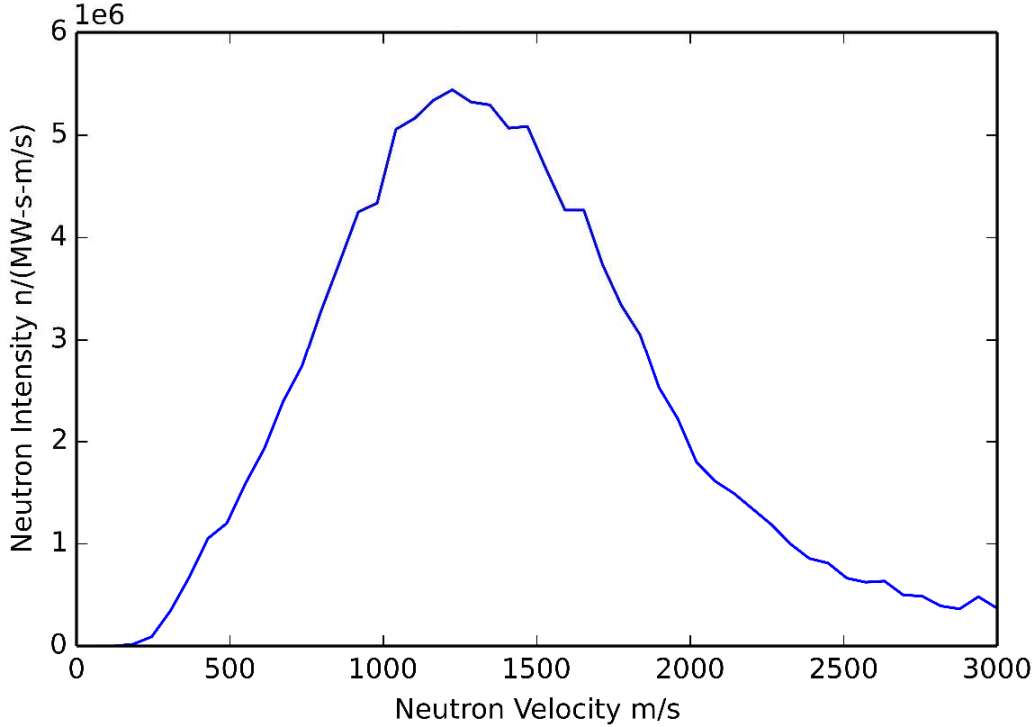}
\caption{Normalized neutron velocity spectrum $S(v)$ of SNS cold beam in McStas simulation.}\label{fig:2}
\end{figure}

Using the McStas neutron simulation package ~\cite{McStas, Gallmeier} we optimized the layout geometry of the hypothetical regeneration search experiment at one of the typical SNS beams. The simulation included the transport with gravity of the neutron beam from the cold moderator through a simple 12.5-m long collimator with opening aperture $10 \times 12$ cm$^2$ providing beam divergence of $< \pm 5$ mrad. Simulated SNS cold beam flux $S(v)$ can be described by the Maxwellian distribution with equivalent temperature $T = 64$K. The integrated neutron intensity was found to be $N_0 = 6\times 10^{9}$ neutrons/sec/MW in the velocity range 200-2000 m/s (wavelength $> 2$ angstroms). Slow neutrons in the spectrum are more valuable for the $n \rightarrow n'$ observation since they could spend more time in the flight tube; however, the effect of gravity causes a lower limit on this velocity.  

We assumed (see Figure 1) a gaseous $^3He$ detector (A) will be operating in current integration mode with typical efficiency 1\% upstream of the first flight tube. Such a detector is common equipment for monitoring the intensity of cold beams where instabilities at the percent level can be expected. With initial beam intensity $R_0$, this detector (A) will measure the current integrated over 1 second that is proportional to $\sim 0.01\times R_0$. 

After the first flight tube, the detector (B) with $\sim 100~\%$  
efficiency can be used. This is a $^3He$ current-integrating counter measuring the full beam intensity including the small disappearance effect. 
It was demonstrated e.g. in ~\cite{Szymanski} that an ionization chamber operating in current integrating mode provides statistical accuracy of intensity measurement comparable with corresponding square root of counting number intensity. Thus, we assume that measurements of the ratio of integrated currents in (B) and (A) counters ${I_B(t)} / {I_A(t)}$ could be used to detect the disappearance effect with appropriate statistical error over the full range of proposed field configurations.

At the end of the second flight tube, the $^3He$ counter (C), operating in the proportional pulsed mode, can be used for counting the regenerated neutrons with an efficiency close to 100 \%. This counter should be well shielded to provide the lowest background counting rate possible.  Assuming that counter (C) will have a transversal size $50 \times 50$~cm$^2$ we found from the comparison with similar existing devices that the achievable background rate can be lower than $n_{bkgr} = 0.1$ counts/sec. We will assume this rate for the further simulations and estimates.  

Efficiency of counter (A) at the level $\lesssim 1\%$ will provide statistically sufficient monitoring accuracy for the regeneration measurements where counting rates are low. However, for disappearance measurements the $1\%$ monitoring efficiency will dominate the statistical accuracy in the determination of ratio of ${I_B(t)} / {I_A(t)}$, thus increasing the statistical error of measurement of disappearance rate by factor of 10. With a gaseous $^3He$ monitoring counter, it will be possible to vary the efficiency by changing the partial pressure of the $^3He$ in gas. The cross section of the detection reaction $n+^3He\to^1H+^3H$ depends on the neutron velocity as $1/v$. Therefore, the large efficiency of the monitoring counter will lead to the absorption of the slowest neutrons that contribute more significantly to the sensitivity of the disappearance and even more so to the regeneration of neutrons. Spectrum efficiency can be calculated and the ratio ${I_B(t)} / {I_A(t)}$ can be optimized in terms of the statistical monitoring error. As was found in our calculations for the SNS spectrum, the optimum can be achieved at the efficiency of $\sim 30\%$. In further estimates, we will consider this efficiency for the measurement of disappearance effect although, for regeneration measurement, such high 
efficiency of monitoring will severely impair the sensitivity by reducing the regeneration rate by a factor of 3. 

An alternative scheme of intensity monitoring for the disappearance measurement can be considered with counter (A) keeping a low efficiency. In this scheme, real time is shared 50\% between measurement of neutron intensity by absorption counter (B) at the given value of magnetic field and monitoring measurement at a null magnetic field. We assume that switching between different values of magnetic field will take a short amount of time as compared to the time $T_M$ of measurement with either field. Thus, such alternation can have 
frequency as high as 1 Hz. For zero magnetic field B, the existing limit for $n \rightarrow n'$ disappearance oscillation time was set in UCN experiments as $\tau > 414 s$ \cite{Serebrov1,Olive:2016xmw}. Thus, the zero value of magnetic field can be used as established no-effect point. For time-sharing monitoring the sharing efficiency of 50\% will be optimal, however at the same time, the rate of regeneration measurement will be reduced by a factor of 2. This scheme can work for the disappearance measurement if the characteristic instability time of the neutron source is essentially larger than $T_M$. This might be more appropriate for the cold beams at the reactors rather than at pulsed spallation sources. Stability for this method of intensity monitoring can be practically studied in a real experiment by continuous operation of counter (B) at the nominal intensity. 

We assume that the magnetic field inside the flight tubes can be generated with a set of 3-dimensional current coils positioned outside the non-magnetic tube. A solenoidal coil can control the magnetic field along the tube axis and two orthogonal ``cos($\theta$)'' coils with wires running along the tube length and with cosine distribution of current on the cylindrical surface can be used to set two uniform dipole components of the field in transversal direction. The cos($\theta$) coils are compact and can be built together with solenoidal coil as concentric cylinders outside the vacuum tube made of thin isolated sheets carrying the current wires. These coils will compensate the local, uniform Earth magnetic field and create a controlled 3-D magnetic field inside the flight tube. If the local, environmental magnetic field will be non-uniform, additional coils can be installed to compensate for these local effects. Alternatively, an outer layer of $\mu$-metal shielding can be installed around the field-shaping coils to reduce the effects of most external field sources. Using active feedback of the currents in respect to the reference calibrated magnetic field sensors, we assume that the magnetic field can be controlled in any direction with magnitude from 0 to 100~$\mu$T with uniformity better than 0.2 $\mu$T. The design and control of the magnetic field will likely be the most technically challenging but feasible problem in the considered experiment. A similar kind of problem for shaping uniform magnetic fields with cos($\theta$) coils was considered and prototyped in the paper ~\cite{PerezGalvan:2011zz} for measurement of the neutron electric dipole moment. In a disappearance/regeneration experiment, the effects of the local magnetic field non-uniformity and fluctuations can result in some deviation from the quasi-free condition as it was shown in Ref. \cite{Davis} for neutron--antineutron oscillation \cite{Phillips}; however, these effects can be kept under control. 

\section{Disappearance Measurement}

The magnitude of the magnetic field during the experiment will be varied in steps of $\Delta B$ that should be chosen small enough as compared to the width of the resonance peak. The width of the disappearance resonance in units of $\Delta B$ is determined by the probability function $P_B(t_1)$ with equations (1) and (2) dependent on the time-of-flight $t_1$ through the counter path length $L_1$. Integrated with the given spectrum of neutron velocities $S(v)$ results in a resonance width in units of $\Delta B$ independent of the magnetic field $B$. Suppose that the initial search scan will be performed in the range of laboratory magnetic field  0 to $\pm 20 \mu$T corresponding to the controversial region detected in the aforementioned UCN experiments. If the uniformity of the magnetic field in the neutron flight path will be e.g. $\lesssim 1\%$, $\Delta B$ can be chosen as 0.2$\mu$T$=2$mG.
Measurements with different values of magnetic field $B$ will be statistically independent, and thus, might contribute to the significance of the signal if the latter will be seen as a resonance in either disappearance or regeneration mode. In case of our example SNS spectrum, we have defined the width of resonance peak
or the ``search window" for disappearance search as 35 consequent measurements separated by step of 0.2$\mu$T and for regeneration search (see below) as 13 measurements. 

The experiment for a neutron disappearance/regeneration search is a simple counting experiment to be performed with 3 counters (see Figure 1): (A) monitoring, (B) absorption, and (C) regeneration counters as described above. Counters (A) and (B), due to high detection rate, are operating in the current-integrating mode and counter (C) in pulse-counting mode.  Let us assume that a measurement with a value of magnetic field $\bf{B}$ will last for time $T$ that can be e.g. 3600 seconds (1 hour). 
By integrating the velocity spectrum $S(v)$, we can define the total initial beam rate $R_0$ or the intensity; see eq. (5). To compensate for the possible time variation of the beam intensity, the counting rate of the absorption counter (B) $R_B$ that is measuring the disappearance effect will be divided by the rate $R_A$ of the monitoring counter (A) that has efficiency $\epsilon$. Average detection rates in (A) are: $R_A=\epsilon \cdot R_0$ and $R_A + R_B = R_0$. Since the absorption cross section in $^3He$ counter is proportional to $1/v$, the monitoring counter (A) will deplete the spectrum of neutrons with low velocities and the probability of disappearance in eq. (1) and (2) will be reduced as some function of the monitoring efficiency. This function for given the velocity spectrum was calculated by direct spectrum integration and approximated as $(1-1.7\epsilon+0.76{\epsilon}^2)$. It includes reduction of the counting rate in counter (B) as well as reduction due to modification of velocity spectrum due to monitoring. We assume that the absorption counter (B) is thick enough to provide total absorption of neutrons within the range of velocities of interest.
Thus, the effect of disappearance (beam intensity measured by counter (B) modified by the disappearance effect and corrected by the intensity monitor), $D$, and the corresponding statistical error $\sigma_D$ of measurement in counter (B) becomes 
\begin{equation}
D = R_0T(1-\epsilon) - R_0T\langle P(t_1) \rangle (1-1.7\epsilon+0.76{\epsilon}^2);~~~~~ \sigma_D=\sqrt{\frac{R_0T(1-\epsilon{})}{\epsilon{}}},
\end{equation}
where $\langle P(t_1) \rangle $ is the average value of probability integrated over the velocity spectrum.  Figure 3 below shows that for a simulated example of the disappearance measurement with $\Delta B = 0.2 \mu$T for $\tau = 3$ s and with monitoring efficiency $\epsilon = 1\%$. The larger disappearance peak in this figure at the resonance value has a significance of more than $10 \sigma$ where we are assuming the contribution of only statistical factors. 

According to eq. (3), the optimum value of $\epsilon$ of counter (A) to  maximize the disappearance effect is $\epsilon=30\%$, as mentioned earlier. In this case, the statistical significance of the peak for every value of magnetic field will be a factor 3.7 higher than for $\epsilon=1\%$. The height of the two resonance peaks in Figure 3 is proportional to factors $(1+cos{\beta})$ and $(1-cos{\beta})$ for negative and positive orientations of vertical magnetic field $\bf B$ with unknown angle $\beta$ between vectors 
$\bf B$ and $\bf B'$ (see eq.(1)-(2)). Since both peaks are generated by the same effect of neutron disappearance, in the analysis of peak significance, or the absence thereof, for determination of the exclusion upper limit for oscillation time $\tau$ it is useful to fold two parts of the scan with positive and negative directions of magnetic field B. In this case the folded resonance peak will not depend on the angle $\beta$ and significance of the peak (or exclusion limit) can be determined in terms of the oscillation time $\tau$ only. On another count, the different relative height of two peaks, as in Figure 3, would provide measurement of the angle $\beta$.
As an example, the angle $\beta$ is assumed in Figure 3 to be equal to $3/4\pi$ relative to the vertical up direction. Thus, in the measurement of disappearance both parameters $\tau$ and $\beta$ of equations (1)-(2) can be separately determined.
In the case of the absence of resonance, the exclusion limit for oscillation time $\tau$ can be set independent of the angle $\beta$ from folded measurement.

The effect of disappearance $D$ described by the equation (3) can be simplified for the magnitude of the magnetic field corresponding to the peak value of the resonance if we will neglect the non-resonance terms in equation (2):

\begin{figure}[t]	
	\centering
\includegraphics[width=480pt, trim=40 80 40 80, clip=true]{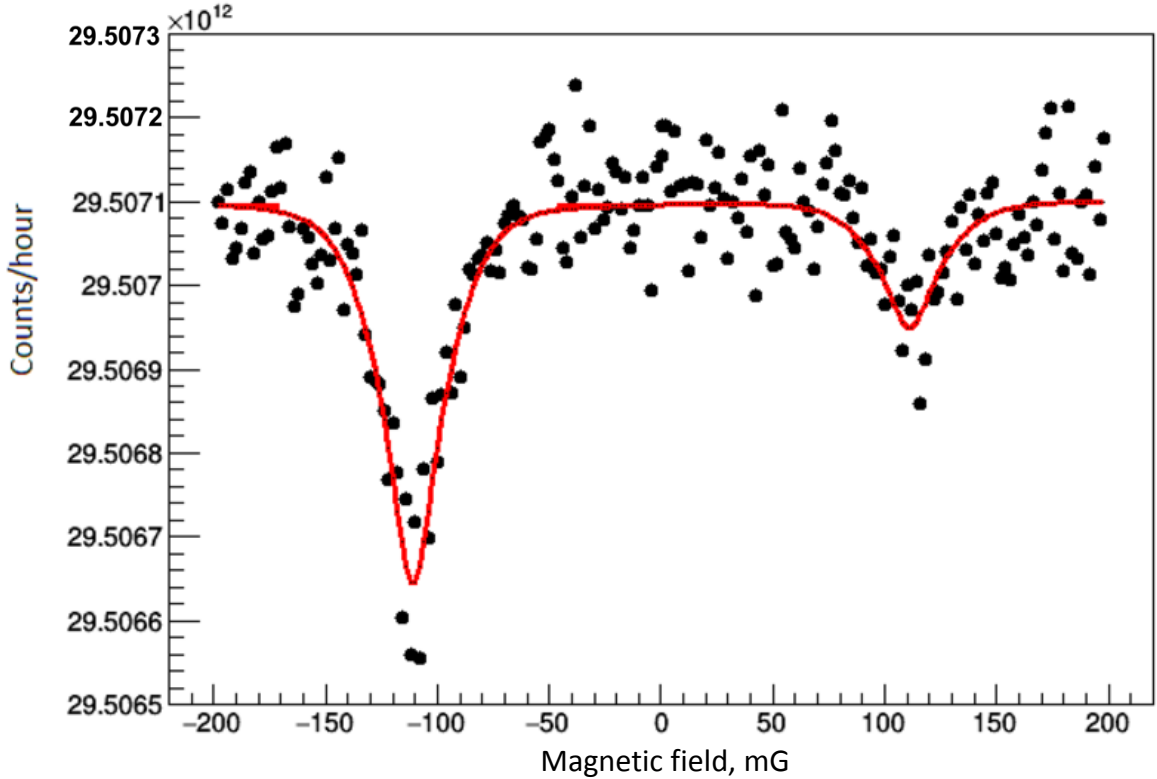}
\caption{Simulated magnetic field scan for disappearance measurement in  absorption counter (B) corrected with $\epsilon = 1\%$ efficiency monitor counter (A)  for $n \rightarrow n'$ oscillation time  $\tau = 3$ s and $cos{\beta}=-0.5$. Unknown magnitude of mirror magnetic field $B'$ was assumed to be equal to 11.1 $\mu$T.}\label{fig:3}
\end{figure}
\begin{equation}
D =R_0T(1-\epsilon) - R_2T(1-1.7\epsilon+0.76{\epsilon}^2) \frac{L^2_1}{2{\tau}^2} (1+cos{\beta}),
\end{equation}
where 
\begin{equation}
R_0 = \int_{200}^{3000} S(v) dv;  ~~~~~
R_2 = \int_{200}^{3000}\frac{S(v)}{v^2} dv 
\end{equation}
\vspace{10pt}

The statistical significance of the resonance peak in the disappearance experiment should be properly determined by a fit procedure with parameters given by eq. (1) and (2). Qualitatively, we can describe the statistical significance as a proper disappearance effect -- negative term in eq. (4) -- in units of statistical error $\sigma_D$, with additional amplification factor $W$ due to contribution of multiple measurements within the width of resonance. The value of the exclusion upper limit for $\tau$ corresponding to this significance level in the disappearance measurement will improve linearly with the flight distance $L_1$ and as $\sqrt [4] T$ of the measurement time $T$ per magnetic field point. Thus, a 10 times longer measurement time $T$ can improve $\tau$ by a factor of 1.78. Increase in the flight path might be more efficient, but can be limited by the constraints of available space at the neutron source facility and also by the effect of gravity. For the detected resonance peak, the determination of fit parameters will include the subtraction of zero-field (no-effect) measurements with a corresponding increase of the statistical error.

For determination of exclusion upper limit of $\tau$ in the disappearance scan, the folded measurements for $\pm B$ can be used. The upper limit of $\tau$ will correspond to the effect (1)-(2) with $\chi^2$ measured value within the sliding resonance ``search window" not exceeding the selected confidence (CL) level of $\chi^2$ fluctuations in the absence of the effect, i.e. in measurements with zero magnetic field. Table 1 below shows 95 \% CL exclusion upper limits for oscillation time $\tau$ for different schemes of monitor efficiency $\epsilon$ and measurement time $T$ as determined in simulations.
\vspace{40pt}

{\raggedright
Table 1. Possible disappearance exclusion upper limits for $+B$ and $-B$ folded
measurements independent of unknown cos$\beta{}$.
\vspace{5pt}
}

{\raggedright
	\centering
\vspace{3pt} \noindent
\begin{tabular}{|p{140pt}|p{130pt}|p{130pt}|}
\hline
\parbox{141pt}{\centering 
T, hours per measurement
} & \parbox{141pt}{\centering 
Monitor efficiency $\epsilon$
} & \parbox{141pt}{\centering 
95\% CL $\tau{}$ limit, s
} \\
\hline
\parbox{141pt}{\centering 
1
} & \parbox{141pt}{\centering 
0.01
} & \parbox{141pt}{\centering 
7.2
} \\
\hline
\parbox{141pt}{\centering 
10
} & \parbox{141pt}{\centering 
0.01
} & \parbox{141pt}{\centering 
12.9
} \\
\hline
\parbox{141pt}{\centering 
1
} & \parbox{141pt}{\centering 
0.30
} & \parbox{141pt}{\centering 
14.0
} \\
\hline
\parbox{141pt}{\centering 
10
} & \parbox{141pt}{\centering 
0.30
} & \parbox{141pt}{\centering 
24.6
} \\
\hline
\parbox{141pt}{\centering 
1
} & \parbox{141pt}{\centering 
Time sharing
} & \parbox{141pt}{\centering 
16.5
} \\
\hline
\parbox{141pt}{\centering 
10
} & \parbox{141pt}{\centering 
Time sharing
} & \parbox{141pt}{\centering 
29.2
} \\
\hline
\end{tabular}
\vspace{15pt}

}

\section{Regeneration Measurement}

Several discussions of the previous section will be also relevant for regeneration measurements. For the regeneration process, the width of the resonance peak follows from the probability product of $P_B(t_1) \cdot P_B(t_2)$ and is significantly narrower than for disappearance. The time $t_1$ and $t_2$ are the neutron flight time for the path $L_1$ and $L_2$ correspondingly. For regeneration measurements we defined the ``search window" as 13 consequent measurements separated by step of 0.2$\mu$T. For these measurements, high efficiency of the monitoring counter (A) is not required and would even significantly impair the transformation probability. Therefore, for regeneration we assume low efficiency ($< 1\%$) of the counter (A) that will not essentially deplete the low-velocity part of the spectrum. However, the counting rate of the counter (A) still will be much higher than any possible rate of the regeneration counter (C). Therefore, monitoring of the beam intensity will not contribute significantly in the statistical error of regeneration measurement. Most essential for observation of a regeneration effect in the $^3He$ counter (C) will be the background due to detection of thermal neutrons. We assume that this irreducible background rate will be $n_{bkgr} = 0.1$ cps which is achievable for a high-efficiency counter of the size $50 \times 50$ cm$^2$. Figure 1 shows the simulated example of magnetic field scan with detection of the regeneration effect. Black points represent counts in every run, and the continuous curve is the sum of the simulated effect and assumed background. The counts in the peak of the resonance correspond to a signal of more than 12$\sigma$
significance. The resonance in the opposite hemisphere of $\bf B$ is statistically insignificant.

\begin{figure}	[t]
	\centering
\includegraphics[width=480pt, trim=40 90 40 80, clip=true]{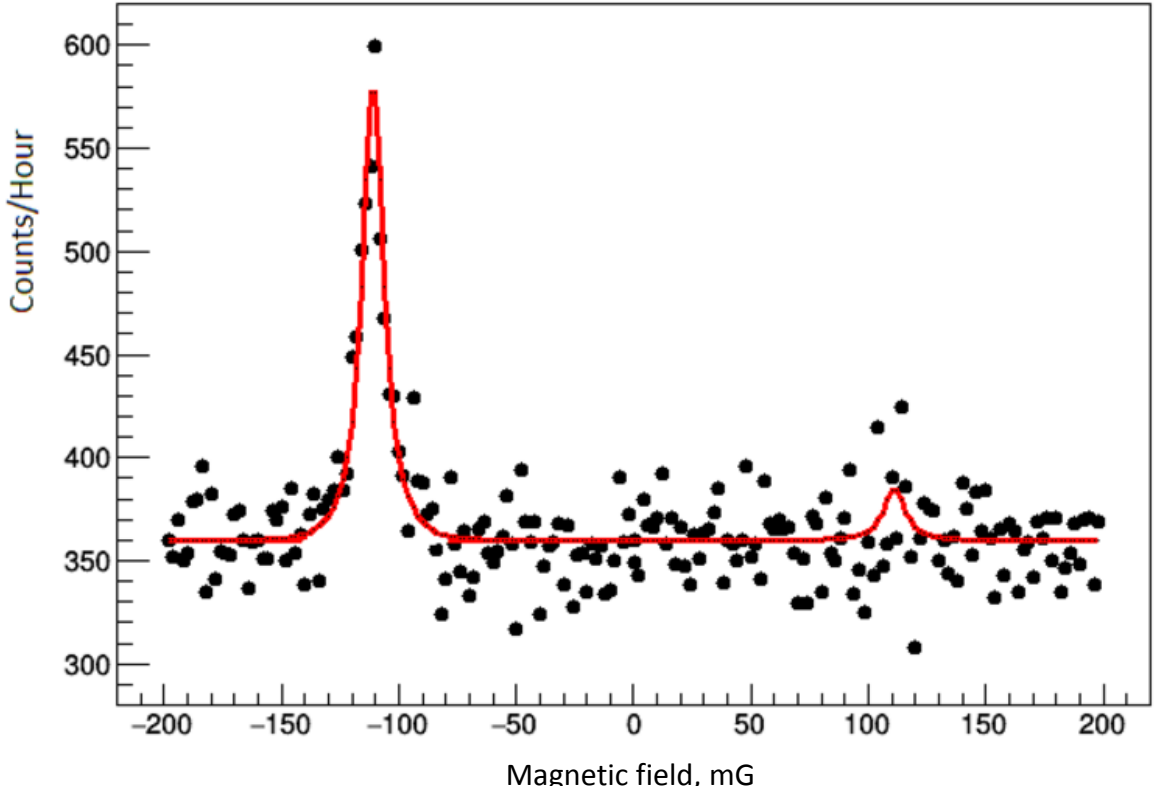}
\caption{Simulated example of magnetic field scan for $n \rightarrow n' \rightarrow n$ regeneration detection. Background rate in $^3He$ counter (C) is assumed to be 0.1 cps,  measurement time $T=1$ hour per point, oscillation time $\tau = 10$ s, and $cos{\beta} = -0.5$.  Unknown magnitude of mirror magnetic field $B'$ was assumed to be equal to 11.1 $\mu$T.}\label{fig:4}
\end{figure}

In the absence of the regeneration effect, e.g. with zero magnetic field and with a sufficient amount of shielding, the background counting rate in counter (C) should be not dependent on the neutron beam intensity. Thus, the effect of regeneration $G$ from eq. (1)-(2) in counts of the counter (C) for the time $T$ and its statistical error $\sigma_G$ at any given value of magnetic field will be correspondingly:
\begin{equation}
G = n_{bkgr}T + \langle P(t_1) \cdot P(t_2) \rangle \cdot (1-\epsilon)R_0T; ~~~~~
\sigma_G=\sqrt{G}
\end{equation}
Here $\langle P(t_1) \cdot P(t_2) \rangle$ is the product of probabilities for flight path $L_1$ and $L_2$ averaged over the velocity spectrum.  As follows from (1), (2) where we can neglect the non-resonance terms, and from eq. (6) the number of counts in the regeneration measurement in counter (C) at resonance peak: 
\begin{equation}
G =n_{bkgr}T+R_4T(1-\epsilon) \frac{L^2_1L^2_2}{4\tau^4}{(1+cos\beta)}^2,
\end{equation}
where 
\begin{equation}
R_4 = \int_{200}^{3000}\frac{S(v)}{v^4} dv 
\end{equation}
\vspace{5pt}

Folding the magnetic field scan measurements for $+B$ and $-B$, according to eq. (7), in case of regeneration measurement does not remove the dependence on $cos{\beta}$, but allows for using both resonance peaks in the treatment of statistical significance. Qualitatively statistical significance 
can be defined for the sliding ``search window" of 13 measurements as the ratio of the regeneration effect from (7) and the statistical error $\sigma_G$ from (6) with an additional amplification factor $W$ due to the contribution of multiple measurements within the width of resonance. As easily seen from this definition, the upper exclusion limit for oscillation time $\tau$ extracted from the scan data will be proportional to $\sqrt{L_1 L_2}$ and $\sqrt [8]T$. If the total length of the experiment will be limited, then the optimal length will be when $L_1 = L_2 = L$ as in Figure 1 and the upper achievable limit for $\tau$ will be proportional to $L$. Table 1 below shows 95 \% CL exclusion upper limit for oscillation time $\tau$ that can be determined from the regeneration measurement for different assumptions for unknown angle $\beta$ and 
for different running times $T$ as obtained in statistical simulations of experiment.

{\center 
\hspace{50pt} 
Table 2. Possible regeneration exclusion limits}
{\center
\begin{tabular}{|p{79pt}|p{79pt}|p{79pt}|p{79pt}|}
\hline
\parbox{79pt}{\centering 
T, hours per measurement
} & \parbox{79pt}{\centering 
Mode
} & \parbox{79pt}{\centering 
Cos $\beta{}$
} & \parbox{79pt}{\centering 
95\% CL $\tau{}$ limit, s
} \\
\hline
\parbox{79pt}{\centering 
1
} & \parbox{79pt}{\centering 
Non-folded
} & \parbox{79pt}{\centering 
-0.5 (example)
} & \parbox{79pt}{\centering 
14
} \\
\hline
\parbox{79pt}{\centering 
10
} & \parbox{79pt}{\centering 
Non-folded
} & \parbox{79pt}{\centering 
-0.5 (example)
} & \parbox{79pt}{\centering 
19
} \\
\hline
\parbox{79pt}{\centering 
1
} & \parbox{79pt}{\centering 
Folded
} & \parbox{79pt}{\centering 
0 (min)
} & \parbox{79pt}{\centering 
13.5
} \\
\hline
\parbox{79pt}{\centering 
10
} & \parbox{79pt}{\centering 
Folded
} & \parbox{79pt}{\centering 
0 (min)
} & \parbox{79pt}{\centering 
18
} \\
\hline
\parbox{79pt}{\centering 
1
} & \parbox{79pt}{\centering 
Folded
} & \parbox{79pt}{\centering 
1 (max)
} & \parbox{79pt}{\centering 
16.1
} \\
\hline
\parbox{79pt}{\centering 
10
} & \parbox{79pt}{\centering 
Folded
} & \parbox{79pt}{\centering 
1 (max)
} & \parbox{79pt}{\centering 
21.4
} \\
\hline
\end{tabular}
\vspace{20pt}

}

\section{Conclusions}

The neutron disappearance/regeneration experiment described above can provide a resolution of the controversy observed in the experiments with UCN traps looking for neutron to mirror neutron disappearance. Both cold neutron beam disappearance and regeneration methods, as described above, will be free of the uncertainties of UCN interactions with material walls. Requirement of monitoring of the intensity of cold neutron beam makes optimization of sensitivity for disappearance and regeneration operation modes different, such that for optimal performance these two measurements should be performed as two separate experiments with different efficiency of the monitoring counter. From Tables 1 and 2 we can see that with minimal anticipated measurement time per scan point as $T = 1$ hour each of these two experiments will require $\sim$ 10 days of beam time. Since the increase of the flight path length can linearly increase the limit of oscillation time $\tau$ it will be advantageous for the disappearance stage of the experiment to increase the total length of the disappearance path to 2$L_1$, or in our example to 30-m. Then, with the SNS example cold neutron beam, monitoring efficiency 30\%, and for 10 days run time in disappearance mode, the limit of $\tau >$ 28 s can be reached. For the regeneration measurement, two flight paths separated by the absorber will be needed and to reach the same level of sensitivity as in the disappearance measurement the overall increase of the experiment length will be required. Increase of the flight path length will make the width of resonance narrower and might increase the requirements for the uniformity of magnetic field. Ultimately, the increase of sensitivity with the length will be counteracted by the gravity effect depleting the low-velocity part of the neutron spectrum. We would like to stress that disappearance and regeneration are two different physics processes; in this paper we discuss them in the context of the theoretical mirror matter model where they have common origin, but these two processes should be studied independently. 

We assumed that each disappearance and regeneration measurement will consist of a systematic magnetic field scan, e.g. from $-20 \mu$T to $+20 \mu$T with a step $0.2\mu$T.  Due to the simplicity of the concept of the experiment, fast feedback from measurements can be available on-line. This will allow, in the case of observation of an anomaly in the counting rates, for the experiment configuration to return to any particular magnetic field setting and re-run this point for a reproducibility check.

Observation of a resonance pattern in the magnetic scan by both methods with consistent oscillation time will be a strong argument for the existence of a new particle, the mirror neutron, and new physics with adequate theoretical description of the effect. Positive detection of a new effect with the oscillation time $\tau$ in the range 10--30 s will facilitate, along side additional experimental efforts, other interesting measurements possibilites. Thus, the vector direction of $\bf{B'}$ mirror magnetic field can be determined, which would confirm the existence of an electromagnetic component for mirror matter. A disappearance and regeneration experiment with magnetic field $\bf{B}$ coinciding in the direction with $\bf{B'}$ will allow detection of the  neutron transformation effect with its maximum unsuppressed magnitude. Since the measurement of a single field configuration with $\bf{B}$ takes a short time on the order of 1 hour, the stability of magnitude and direction of the mirror magnetic field $\bf{B'}$ can be explored giving the insight on the cosmological or terrestrial origin of $\bf{B'}$. At the resonance condition, the beam of mirror neutrons as massive sterile particles penetrating the absorption wall will be formed behind the wall, thus, allowing the study of possible interactions of MM with OM.  

In the absence of neutron disappearance and regeneration by methods considered in this paper, the controversy of existing UCN $n\to n'$ disappearance results will be resolved. It will also exclude, at least at the level of the exclusion limit for $\tau(n\to n')$, the possible contribution of $n\to n'$ transformations to the observed difference of neutron lifetime obtained in appearance and disappearance modes as mentioned e.g. in ~\cite{Wietfeldt:2014rea}.

All simulations in this paper were done using the SNS parameters as an example. The described experiment can be performed at several neutron sources where cold neutron beams of high intensity are available. Neutron sources with colder neutron spectra, with higher beam intensities, or where beam line design can be improved by the focusing of cold neutrons with corresponding increase of the length of the experiment potentially can provide larger sensitivities. In particular, the European Spallation Source (ESS) that is under construction at Lund \cite{ESS} with the Large Beam Port 
envisaged in the project for the neutron-antineutron search experiment, can provide sensitivity for an oscillation time of more than 50 seconds. This is due to higher intensity and flight times that can be accomodated at that facility.
In this way, if the $n-n'$ oscillation time happens to be around 10 seconds, the ESS in fact could become a factory of mirror neutrons.

\section{Acknowledgments}

We are grateful to ORNL colleagues Leah Broussard, Franz Gallmeier, and Erik Iverson, and also to Christopher Crawford 
from the University of Kentucky, Chen-Yu Liu and William M. Snow from the Indiana University for useful discussions. We also appreciate an interest and encouragement of our colleagues from the University of Tennessee Yuri Efremenko, Geoffrey Greene, Lawrence Heilbronn, Caleb Redding, Arthur Ruggles, and Lawrence Townsend. This work was supported in part by US DOE Grant DE-SC0014558 and UTK ORU program.

\end{document}